\documentclass[aps,showpacs,showkeys,amssymb]{revtex4}

\begin{document}

\title{Some universal relations between the gap and thermodynamic
       functions plausible for various models of superconductors}

\author{Ryszard Gonczarek}
\author{Mateusz Krzyzosiak}

\affiliation{Institute of Physics, Wroc{\l}aw University of Technology,
         Wybrze\.{z}e Wyspia\'{n}skiego 27, 50-370 Wroc{\l}aw, Poland}

\date{\today}

%%%%%%%%%%%%%%%%%%%%%%%%%%%%%%%%%%%%%%%%%%%%%%%%%%%%%%%%%%%%%%%%%%%%%%%
% Abstract
%%%%%%%%%%%%%%%%%%%%%%%%%%%%%%%%%%%%%%%%%%%%%%%%%%%%%%%%%%%%%%%%%%%%%%%

\begin{abstract}
  It is proven that there exist some universal relations between the energy
  gap and the differences of the thermodynamic potential, entropy,
  specific heat and critical magnetic field for many
  two- and three-dimensional models of
  superconductivity with spin-singlet Cooper pairs forming
  \textit{s} or \textit{d} or \textit{g} etc. states. The obtained
  formulae make it possible to derive thermodynamic functions and,
  in particular, the superconducting specific heat and the
  critical magnetic induction in the whole
  temperature range $0\leq T\leq T_{c}$ employing the form of
  $\Delta(T)$ only.
  The inverse formula allowing us to find $\Delta(T)$, when the
  temperature dependence of the  specific heat difference is known,
  is also presented.
  The results are referred to some obtained within the McMillan
  formalism, and some remarks on an application of the
  present formulae to the {\textit{t--J}} model are given.
\end{abstract}

%%%%%%%%%%%%%%%%%%%%%%%%%%%%%%%%%%%%%%%%%%%%%%%%%%%%%%%%%%%%%%%%%%%%%%%
% PACS numbers and key words
%%%%%%%%%%%%%%%%%%%%%%%%%%%%%%%%%%%%%%%%%%%%%%%%%%%%%%%%%%%%%%%%%%%%%%%

\pacs{74.20.-z,74.25.Bt} \keywords{energy gap amplitude,
thermodynamic functions, specific heat, spin-singlet states}

\maketitle

%%%%%%%%%%%%%%%%%%%%%%%%%%%%%%%%%%%%%%%%%%%%%%%%%%%%%%%%%%%%%%%%%%%%%%%
\section{Introduction}
%%%%%%%%%%%%%%%%%%%%%%%%%%%%%%%%%%%%%%%%%%%%%%%%%%%%%%%%%%%%%%%%%%%%%%%

Nowadays, there exist several approaches which are able to explain
properties of high-$T_{c}$ and other new-generation superconductors.
In most of these approaches only one common microscopic mechanism
responsible for superconductivity in all high-$T_{c}$
superconductors is presented. However, this mechanism has not been
univocally identified so far, although particularly preferred is
the coupling through electron-electron interaction \cite{pap1},
or spin exchange \cite{pap2,pap3}, as well as a modified
phonon-mediated established by Anderson
\cite{pap7,pap8,pap9}, based on the concept
of the resonating valence bond states.
On the other hand, in all these approaches the pairing interaction is
modeled by means of the exchange-mediated coupling which leads to the
formation of Cooper pairs. Then, in order to explain the enhancement
of the transition temperature, one can employ the Van Hove scenario
\cite{pap10,pap11,pap12,pap13,pap14,pap15,pap16,pap17,pap18,pap19}.
Recently it has been shown \cite{pap20}
that starting in the real space by setting a Hamiltonian, which
contains atomic or itinerant electrons energy, all hopping and exchange
integrals and all two-site interactions, and then solving the model
Hamiltonian within the mean-field approximation, one can transfer the
problem into the reciprocal space. Performing a specific
conformal transformation of the reciprocal space, one can reduce the
problem to the BCS-type one, where all elements of the symmetry and the
dispersion relation are gathered in the Jacobian of the conformal
transformation. Such Jacobian describes a specific scalar field of the
density of states which is imposed on the BCS-type isotropic reciprocal
space, and which can contain some singularities and nodes. The Jacobian
appears in all equations and its mean value can be identified with the
density of states. Therefore, the proposed formalism \cite{pap20} is an
extension of the Van Hove scenario and can be applied for
non-\textit{s}-wave symmetry systems.

In the present paper we demonstrate that for a wide class of two-
and three-dimensional models of superconductors with electrons or
holes as carriers, and spin-singlet Cooper pairs forming a
\textit{s} or \textit{d} or \textit{g} etc. state, there exist some
universal relations between the energy gap and the thermodynamic
potential, entropy and specific heat differences defined between
the superconducting and the normal phase. Moreover, we show that
the obtained formulae allow us to derive the critical magnetic
induction and its derivatives. We also discuss the case when
strong-coupling electron-phonon effects are included, and we give
remarks on an application of the present formulae to
a relevant model of superconductivity i.e. the {\textit{t--J}} model.

%%%%%%%%%%%%%%%%%%%%%%%%%%%%%%%%%%%%%%%%%%%%%%%%%%%%%%%%%%%%%%%%%%%%%%%
\section{Gap equation}
%%%%%%%%%%%%%%%%%%%%%%%%%%%%%%%%%%%%%%%%%%%%%%%%%%%%%%%%%%%%%%%%%%%%%%%

Although the gap equation can be obtained in a self-consistent
manner (e.g. in the Green function formalism) we
emphasize that it defines the energy gap which minimizes the free
energy, so it can be found from the free energy variation. Because
in all known approaches the one-particle energy of eigenstates is
modified by the term $E_{\bf k}=\sqrt{(\epsilon_{\bf k}-\mu)
^{2}+|\Delta_{\bf k}|^{2}}$, we postulate that the gap equation,
which can be obtained in various approaches,
formulated for two- or three-dimensional systems can be always
written in the following general form
\begin{equation}\label{1a}
  \Delta_{\bf k}=\frac{1}{N}\sum_{\bf k^{\prime}}V_{\bf
  k,k^{\prime}}\frac{\Delta_{\bf k^{\prime}}}{E_{\bf k^{\prime}}}
  f\left(\frac{\epsilon_{\bf k^{\prime}}-\mu}{T},
  \frac{|\Delta_{\bf k^{\prime}}|}{T}\right),
\end{equation}
where $V_{\bf k,k^{\prime}}$ is a pairing interaction, $\mu$
denotes the chemical potential and $f$ is a certain analytic
function properties of which are defined below. This equation
should become complete with another self-consistent equation
\begin{equation}\label{1b}
  2n=\frac{1}{N}\sum_{\bf k}\left[1-\frac{\epsilon_{\bf
  k}-\mu}{E_{\bf k}}f\left(\frac{\epsilon_{\bf k}
  -\mu}{T},\frac{|\Delta_{\bf k}|}{T}\right)\right],
\end{equation}
which gives the chemical potential $\mu$ as a function of the filling
of the conduction band $n$, for the normal metallic phase.
Eq.~(\ref{1a}) after performing the conformal transformation of
the momentum space \protect\cite{pap20} turns into the form
\begin {equation}\label{1}
  \Delta(\xi,\vartheta,\varphi)=\frac{1}{4\pi^{d-1}}\left<\int_{-\omega_{d}
  }^{\omega_{u}}V(\xi,\xi^{\prime},\vartheta,\vartheta^{\prime},
  \varphi,\varphi^{\prime})\frac{J(\xi^{\prime},\vartheta^{\prime},
  \varphi^{\prime})\Delta(\xi^{\prime},\vartheta^{\prime},\varphi^{\prime})}
  {\sqrt{(\xi^{\prime}-\mu^{\ast})^{2}+|\Delta(\xi^{\prime},\vartheta^{\prime},
  \varphi^{\prime})|^{2}}}
  f\left(\frac{\xi^{\prime}-\mu^{\ast}}{T},\frac{|\Delta(\xi^{\prime},
  \vartheta^{\prime},\varphi^{\prime})|}{T}\right)\;d\xi^{\prime}\right>,
\end{equation}
where $d=2$, or $3$ is the dimension of the reciprocal space,
$\mu^{\ast}$ defines the chemical potential measured from the
Fermi level, so $\mu=\epsilon_{F}+\mu^{\ast}$ and $\xi$ measures
the energy from the Fermi level. $\omega_{u}$ and $\omega_{d}$
mark the limits of integration which are the Debye energy
$\omega_{D}$ for sufficiently wide conduction bands and one of
them or both  must be replaced by conduction band borders for
partially-filled bands of width $2\omega_{b}=\omega_{u}+\omega_{d}$,
when $\omega_{b}<\omega_{D}$. Then $\omega_{u}$ and $\omega_{d}$
can be identified with non-symmetric cut-off parameters of the pairing
interaction, and for sufficiently narrow conduction bands all
particles get paired. When the system is almost but less than
half-filling and $n=1-\delta$, where $\delta$ is small, we have
\begin{equation}\label{1c}
  \omega_{d}=(1-\delta)\omega_{b} \qquad {\rm and} \qquad
  \omega_{u}=(1+\delta)\omega_{b}.
\end{equation}
So, for some superconducting systems the Debye energy $\omega_{D}$
must be replaced by the half band width $\omega_{b}$ after
including additional factors connected with the filling of the
band $n$. Therefore, we can state that the range of the pairing
interaction is given by the energy $\omega$ which represents
$\omega_{D}$ or $\omega_{b}$ according to Eq.~(\ref{1c}) and
hereafter we apply the following notation
\begin{equation}\label{1d}
  \omega_{d}=n\omega \qquad {\rm and} \qquad
  \omega_{u}=(2-n)\omega,
\end{equation}
where for the case of a symmetric pairing interaction as
e.g. for the BCS-type models one has to put $n=1$ and
$\omega=\omega_{D}$. Moreover, $<\ldots>$ denotes the average over
spherical angles $\vartheta^{\prime}$ and $\varphi^{\prime}$ for
the three-dimensional case and over the angle $\varphi^{\prime}$ only
for the two-dimensional case, where
$\vartheta^{\prime}=\vartheta=\frac{\pi}{2}$ are consistently
omitted. $J(\xi,\vartheta,\varphi)$ is the Jacobian of the
employed conformal transformation, and the pairing interaction for
pure $s,d,\ldots$ pairing ($l=0,2,\ldots$) can be taken in the
form
\begin{equation}\label{2}
  V(\xi,\xi^{\prime},\vartheta,\vartheta^{\prime},\varphi,\varphi^{\prime})
  =V_{l}(\xi,\xi^{\prime})\sum_{m=-l}^{l}Y_{lm}(\vartheta,\varphi)
  Y_{lm}^{\ast}(\vartheta^{\prime},\varphi^{\prime}),
\end{equation}
where $Y_{lm}(\vartheta,\varphi)$ are the spherical harmonics.
Supposing that $V_{l}(\xi,\xi^{\prime})$ contains only the
attractive part of interaction, we can replace it by
$g_{l}v_{l}(\xi)v_{l}(\xi^{\prime})$. Then, for systems of
strongly screened electrons we can assume $v_{l}(\xi)=1$, whereas
for a strong electronic repulsion $v_{l}(\xi)$ must be an odd
function of quasiparticle energy \protect\cite{pap21,pap22,pap23}.
$g_{l}$ is a coupling constant and its effective value should be
obtained in a comprehensive renormalizing procedure including e.g.
strong coupling effects and Coulomb repulsion of electrons forming
a Cooper pair \protect\cite{pap24}. In order to be in accordance
with the specified conditions we have to allow for the energy gap
to be of the following form
\begin{equation}\label{3}
  \Delta(\xi,\vartheta,\varphi)=\Delta(T)\;D_{l}(\vartheta,\varphi)\;v_{l}(\xi),
\end{equation}
where $\Delta(T)$ is the amplitude of the energy gap depending on
temperature only, and
\begin{equation}\label{4}
  D_{l}(\vartheta,\varphi)=\sum_{m=-l}^{l}d_{m}(l)Y_{lm}(\vartheta,\varphi)
\end{equation}
expresses the spatial structure of the energy gap on equi-energy
surfaces. $d_{m}(l)$ are normalization constants chosen such, that
the condition $<|D_{l}(\vartheta,\varphi)|^{2}>=1$ is satisfied.
Moreover, since for the most models under consideration the
function $f$ is of the form
\begin{equation}\label{5}
  f\left(\frac{\xi-\mu^{\ast}}{T},\frac{|\Delta(\xi,
  \vartheta,\varphi)|}{T}\right)
  =\tanh\left(\frac{\sqrt{(\xi-\mu^{\ast})^{2}+|\Delta(\xi,
  \vartheta,\varphi)|^{2}}}{2T}\right),
\end{equation}
as a consequence of the relation $\tanh(\frac{x}{2})=1-2n(x)$, where
$n(x)$ is the mean occupation number, we assume that in the limit
$T=0$ it reduces to $1$, whereas for
$\Delta(\xi,\vartheta,\varphi)=0$ we have $f(x,0)\sim x$ when
$x\rightarrow 0$. After employing Eq.~(\ref{3}) and including
Eq.~(\ref{4}) the gap equation (\ref{1}) can be reduced to the form
\begin {equation}\label{6}
  \frac{1}{g_{l}}=\frac{1}{4\pi^{d-1}}\left<\int_{-n\omega}
  ^{(2-n)\omega}\frac{J(\xi,\vartheta,\varphi) v^{2}_{l}(\xi)
  |D_{l}(\vartheta,\varphi)|^{2}}
  {\sqrt{(\xi-\mu^{\ast})^{2}+ v^{2}_{l}(\xi)
  |D_{l}(\vartheta,\varphi)|^{2}\Delta^{2}(T)}}
  f\left(\frac{\xi-\mu^{\ast}}{T},\frac{v_{l}(\xi)
  |D_{l}(\vartheta,\varphi)|\Delta(T)}{T}\right)\;d\xi\right>.
\end{equation}
The obtained formula allows us to derive $\Delta(T)$ for a given
structure of the energy gap when the Jacobian, $v_{l}(\xi)$ and
the function $f$ are known.

Let us focus now on the cases when
$v_{l}(\xi)\equiv 1$ for all $\xi$ or $v_{l}(\xi)\cong$ const for
$|\xi-\mu^{\ast}|$ less than a few $T_{c}$, then we denote it by
$v_{l}$, and when the conduction band is wide enough to imply
$\Delta(0)\ll \omega_{b}$, and when the chemical potential $\mu$
is independent of temperature and equal to its value for a
normal metal at $T=0$, and $\mu^{\ast}$ can be
omitted \protect\cite{pap24a}.
Then the above conditions allow us to ignore
changes of $v_{l}(\xi)$ for other values of $|\xi-\mu^{\ast}|$ and
leave Eq.~(\ref{1b}) out of account \protect\cite{pap24b}. Hence,
in the limit $T=0$ we have $f=1$ and integrating by parts over
$\xi$ the rhs of Eq.~(\ref{6}) we obtain
\begin {eqnarray}\label{7}
  \frac{4\pi^{d-1}}{g_{l}v_{l}^{2}}&=&\left<J((2-n)\omega,\vartheta,
  \varphi)|D_{l}(\vartheta,\varphi)|^{2}\right>\ln[2(2-n)\omega]%\nonumber\\
  +\left<J(-n\omega,\vartheta,\varphi)|D_{l}(\vartheta,\varphi)|^{2}
  \right>\ln(2n\omega)\nonumber\\
  &&-2\left<J(0,\vartheta,\varphi)
  |D_{l}(\vartheta,\varphi)|^{2}\ln (v_{l}|D_{l}(\vartheta,\varphi)|)\right>
  -\left<[J((2-n)\omega,\vartheta,
  \varphi)+J(-n\omega,\vartheta,\varphi)]|D_{l}(\vartheta,\varphi)|^{2}
  \right>\ln\Delta(0)\nonumber\\
  &&- \left<|D_{l}(\vartheta,\varphi)|^{2}\,\int_{-\frac{n\omega}
  {\Delta(0)}}^{\frac{(2-n)\omega}{\Delta(0)}}\ln[u+\sqrt{u^{2}+
  v_{l}^{2}|D_{l}(\vartheta,\varphi)|^{2}}]
  \frac{\partial}{\partial u}
  J(u\Delta(0),\vartheta,\varphi)\;du\right>.
\end{eqnarray}
If the last integral can be considered as constant to the
postulated accuracy, Eq.~(\ref{7}) can be reduced to the form
\begin{equation}\label{8}
  \frac{1}{g_{l}}=N(n)\ln\frac{A}{\Delta(0)},
\end{equation}
where $A$ is positive and may be considered as independent of
$\Delta(0)$, and
\begin{equation}\label{9}
  N(n)=\frac{v_{l}^{2}}{4\pi^{d-1}}\left<[J((2-n)\omega),\vartheta,\varphi)
  +J(-n\omega,\vartheta,\varphi)]|D_{l}(\vartheta,\varphi)|^{2}\right>
\end{equation}
can be understood as a specifically defined mean value of the
density of states \protect\cite{pap20}. Because the obtained
relations hold for an arbitrary value $\Delta(T)$, we have
\begin{equation}\label{10}
  N(n)\ln\frac{A}{\Delta(0)}=
  \frac{1}{4\pi^{d-1}}
  \left<\int_{-n\omega}^{(2-n)\omega}
  \frac{J(\xi,\vartheta,\varphi)v^{2}_{l}(\xi)|D_{l}(\vartheta,\varphi)|^{2}}
  {\sqrt{(\xi-\mu)^{2}+v^{2}_{l}(\xi)|D_{l}(\vartheta,\varphi)|^{2}\Delta^{2}(T)}}
  \;d\xi\right>.
\end{equation}
Hence, subtracting Eq.~(\ref{10}) from Eq.~(\ref{6}) we can
transform the gap equation to the form
\begin{equation}\label{11}
  \frac{1}{\lambda}=\ln\frac{A}{\Delta(T)}+F\left(n,
  \frac{\omega}{T},\frac{\Delta(T)}{T}\right),
\end{equation}
where $\lambda=g_{l}N(n)$ is the dimensionless coupling parameter,
$F$ is a functional of $\Delta(T)$ obtained after integration over
$\xi$ and the angles $\vartheta$ and $\varphi$ for a given form of
the Jacobian, the energy gap structure and the function $f$.

Let us investigate the form of the gap equation (\ref{6}) in the
limit $T=T_{c}$ when $\Delta(T_{c})=0$. Integrating by parts over
$\xi$ and taking into account the above assumptions we can
transform it to the following form
\begin {eqnarray}\label{12}
  \frac{4\pi^{d-1}N(n)}{\lambda v_{l}^{2}}&=&\left<J((2-n)\omega,
  \vartheta,\varphi)|D_{l}(\vartheta,\varphi)|^{2}\right>f
  \left(\frac{(2-n)\omega}
  {T_{c}}\right)\ln[(2-n)\omega]\nonumber\\
  &&+\left<J(-n\omega,\vartheta,\varphi)|D_{l}(\vartheta,\varphi)|^{2}
  \right>f\left(-n\frac{\omega}{T_{c}}\right)\ln(n\omega)\nonumber\\
  &&-\left<\left[f\left((2-n)\frac{\omega}{T_{c}}\right)J((2-n)\omega,\vartheta,
  \varphi)
  +f\left(-n\frac{\omega}{T_{c}}\right)J(-n\omega,\vartheta,
  \varphi)\right]|D_{l}(\vartheta,\varphi)|^{2}\right>\ln T_{c}\nonumber\\
  &&- \int^{(2-n)\frac{\omega}{T_{c}}}_{0}\ln u \frac{\partial}{\partial u}
  \left[f(u)\left<J(uT_{c},\vartheta,\varphi)|D_{l}(\vartheta,\varphi)|
  ^{2}\right>\right]du\nonumber\\
  &&- \int^{\frac{n\omega}{T_{c}}}_{0}\ln u \frac{\partial}{\partial u}
  \left[f(-u)\left<J(-uT_{c},\vartheta,\varphi)|D_{l}(\vartheta,\varphi)|
  ^{2}\right>\right]du,
\end{eqnarray}
which can be written as
\begin{equation}\label{13}
  \frac{1}{\lambda}=\frac{1}{b}\ln\frac{B}{T_{c}},
\end{equation}
where $B$ and $b$ are positive and, due to the postulated
accuracy, may be considered as independent of $T_{c}$, and
\begin{equation}\label{14}
  b=\frac{\left<[J((2-n)\omega,\vartheta,\varphi)
  +J(-n\omega,\vartheta,\varphi)]|D_{l}(\vartheta,\varphi)|^{2}\right>}
  {\left<\left[f\left((2-n)\frac{\omega}{T_{c}}\right)J((2-n)\omega,\vartheta,
  \varphi)+f\left(-n\frac{\omega}{T_{c}}\right)J(-n\omega,\vartheta,
  \varphi)\right]|D_{l}(\vartheta,\varphi)|^{2}\right>},
\end{equation}
is independent of $v_{l}$. Note that the characteristic
ratio
\begin{equation}\label{15}
  \frac{\Delta(0)}{T_{c}}=\frac{A}{B}
  \exp\left(\frac{b-1}{\lambda}\right),
\end{equation}
unlike in the BCS theory, is not a universal constant.

%%%%%%%%%%%%%%%%%%%%%%%%%%%%%%%%%%%%%%%%%%%%%%%%%%%%%%%%%%%%%%%%%%%%%%%
\section{Energy gap as a function of the coupling parameter}
%%%%%%%%%%%%%%%%%%%%%%%%%%%%%%%%%%%%%%%%%%%%%%%%%%%%%%%%%%%%%%%%%%%%%%%

The difference between the thermodynamic potential of the
superconducting and the normal phase, which is equal to the free energy
difference when the chemical potential of both phases is identical, can
be derived from the relation
\begin{equation}\label{16}
  \Delta\Omega(T)=-N(n)\int_{0}^{\lambda}
  \frac{\Delta(T)}{(\lambda^{\prime})^{2}}d\lambda^{\prime},
\end{equation}
where we have taken into account that $g_{l}=\lambda/N(n)$. The
formulas (\ref{8}) and (\ref{13}) allow us to express $\Delta(0)$ and
$T_{c}$ by means of an arbitrary $\lambda^{\prime}$
varying from $0$ to $\lambda$ in the following forms
\begin{equation}\label{17}
  \Delta_{\lambda^{\prime}}(0)=
  A\exp\left(-\frac{1}{\lambda^{\prime}}\right)
  \qquad {\rm{and}} \qquad T_{\lambda^{\prime}}=B\exp\left(-\frac{b}
  {\lambda^{\prime}}\right),
\end{equation}
where, to the assumed accuracy, we can also put $A=A_{1}\omega$
and $B=B_{1}\omega$, where $A_{1}$ and $B_{1}$ are real numbers.
In order to keep the standard notation, hereafter we assume that
$\Delta_{\lambda}(0)\equiv\Delta(0)$ and $T_{\lambda}\equiv T_{c}$,
and moreover $\Delta_{\lambda}(T)\equiv\Delta(T)$. Note that $T_{c}$ is
a decreasing function of $b$, and for the BCS model $b=1$. Moreover,
after eliminating $\lambda$ from these formulas we can find the
characteristic ratio (\ref{15}) as a function of $\omega$. Then
\begin{equation}\label{17a}
  \frac{\Delta(0)}{T_{c}}=A_{1}B_{1}^{-\frac{1}{b}}
  \left(\frac{\omega}{T_{c}}\right)^{\frac{b-1}{b}}\qquad {\rm{or}}
  \qquad \frac{\Delta(0)}{T_{c}}=A_{1}^{b}B_{1}^{-1}
  \left(\frac{\omega}{\Delta(0)}\right)^{b-1},
\end{equation}
and it becomes an increasing or a decreasing function of $\omega$ when
$b>1$ or $b<1$, respectively.

Employing the formulas (\ref{17}) we can find the thermodynamic
potential difference in the limit $T=0$, and we state that it can be
expressed in the following universal form
\begin{equation}\label{18}
  \Delta\Omega(0)=-\frac{1}{2}N(n)\Delta^{2}(0).
\end{equation}
In order to derive the thermodynamic potential difference for
non-zero temperature we have to express $\Delta_{\lambda^{\prime}}(T)$,
i.e. the solution of the gap equation (\ref{11}) for $\lambda^{\prime}$
varying from $0$ to $\lambda$, by means of $\Delta(T^{\prime})$
where $T^{\prime}$ can change from $0$ to $T_{c}$. Let us suppose
that we can find the energy gap for all values $\lambda^{\prime}$ as
a function of reduced temperature $T/\Delta_{\lambda^{\prime}}(0)$.
Then introducing the reduced gap in the form
\begin{equation}\label{19}
  \Lambda_{\lambda^{\prime}}
  \left(\frac{T}{\Delta_{\lambda^{\prime}}(0)}\right)=
  \frac{\Delta_{\lambda^{\prime}}\left(\frac{T}
  {\Delta_{\lambda^{\prime}}(0)}\right)}{\Delta_{\lambda^{\prime}}(0)},
\end{equation}
where, moreover, $\Lambda_{\lambda}(\ldots)\equiv\Lambda(\ldots)$,
and employing Eq.~(\ref{17}) we can transform Eq.~(\ref{11}) to the
form
\begin{equation}\label{20}
  \ln\Lambda_{\lambda^{\prime}}\left(\frac{T}
  {\Delta_{\lambda^{\prime}}(0)}\right)=F\left(n,\frac{\omega}{T},
  \frac{\Lambda_{\lambda^{\prime}}\left(\frac{T}{\Delta_{\lambda^
  {\prime}}(0)}\right)}{\frac{T}{\Delta_{\lambda^{\prime}}(0)}}\right).
\end{equation}
Since this equation has the same form for all values $\lambda^{\prime}$
from $0$ to $\lambda$ when $T,\;\omega$ and $n$ are fixed, all forms of
$\Lambda_{\lambda^{\prime}}\left(T/\Delta_{\lambda^{\prime}}(0)\right)$
derived from Eq.~(\ref{20}) must be equivalent. They become identical
for temperatures normalized by corresponding critical temperatures
denoted as $\tau$.
This observation allows us to express the reduced gap derived for an
arbitrary $\lambda^{\prime}$ by means of the reduced gap $\Lambda$,
which represents the real energy gap of a superconductor
\begin{equation}\label{21}
  \Lambda_{\lambda^{\prime}}\left(\tau\cdot\frac{T_{\lambda^{\prime}}}
  {\Delta_{\lambda^{\prime}}(0)}\right)=
  \Lambda\left(\tau\cdot\frac{T_{c}}{\Delta(0)}\right).
\end{equation}
Hence, after some transformations we find that the value of an
arbitrary reduced gap $\Lambda_{\lambda^{\prime}}$ at
$T/T_{\lambda^{\prime}}$ can be expressed by means of the
function $\Lambda$ as
\begin{equation}\label{22}
  \Lambda_{\lambda^{\prime}}\left(\frac{T}{T_{\lambda^{\prime}}}\right)
  =\Lambda\left(\frac{T}{T_{c}}\cdot\left(\frac{T_{c}}
  {T_{\lambda^{\prime}}}\right)^{2}\cdot
  \frac{\Delta_{\lambda^{\prime}}(0)}{\Delta(0)}\right),
\end{equation}
and the inclusion of Eqs.~(\ref{17}) and (\ref{19}) results in the
following fundamental relation
\begin{equation}\label{23}
  \Delta_{\lambda^{\prime}}\left(\frac{T}{T_{\lambda^{\prime}}}\right)=
  \Delta\left(\frac{T}{T_{c}}\cdot
  \exp\left[(2b-1)\left(\frac{1}{\lambda^{\prime}}
  -\frac{1}{\lambda}\right)\right]\right)\cdot
  \exp\left[-\left(\frac{1}{\lambda^{\prime}}
  -\frac{1}{\lambda}\right)\right].
\end{equation}
Thus, in order to find the values of the energy gap for
$\lambda^{\prime}<\lambda$ and for a fixed temperature $T$, we have to
know the real energy gap values for all higher temperature values if
$b>\frac{1}{2}$, and for all lower temperature values if
$b<\frac{1}{2}$. For $b=\frac{1}{2}$ the temperature dependence
coincides.

%%%%%%%%%%%%%%%%%%%%%%%%%%%%%%%%%%%%%%%%%%%%%%%%%%%%%%%%%%%%%%%%%%%%%%%
\section{Universal relations}
%%%%%%%%%%%%%%%%%%%%%%%%%%%%%%%%%%%%%%%%%%%%%%%%%%%%%%%%%%%%%%%%%%%%%%%
\label{unirel}

The derivation of the relation (\ref{23}) creates new possibilities
to find the thermodynamic potential difference and its derivatives
with respect to temperature as functionals of $\Delta(T)$.
Substituting Eq.~(\ref{23}) into formula (\ref{16}), after some
calculations, we find the thermodynamic potential for
$b\neq\frac{1}{2}$ in the form
\begin {equation}\label{24}
  \Delta\Omega(T)=-\frac{1}{2b-1}N(n)T^{\frac{2}{2b-1}}
  \int_{T}^{T^{\ast}}\frac{\Delta^{2}(T^{\prime})}{(T^{\prime})^
  {\frac{2}{2b-1}+1}}dT^{\prime},
\end{equation}
where $T^{\ast}=T_{c}$ if $b>\frac{1}{2}$ because $\Delta(T)=0$ for
$T\geq T_{c}$, and $T^{\ast}=0$ if $b<\frac{1}{2}$, whereas for
$b=\frac{1}{2}$ we have
\begin {equation}\label{25}
  \Delta\Omega(T)=-\frac{1}{2}N(n)\Delta^{2}(T),
\end{equation}
and hereafter we do not consider this case. The formula (\ref{24}) can
be also written in an another quite equivalent form
\begin {equation}\label{26}
  \Delta\Omega(T)=-\frac{1}{2b-1}N(n)T^{\frac{2}{2b-1}}
  \int_{T}^{T^{\ast}}
  \frac{\Delta^{2}(T^{\prime})-\Delta^{2}(T)}{(T^{\prime})^
  {\frac{2}{2b-1}+1}}dT^{\prime}
  -\frac{1}{2}N(n)\Delta^{2}(T)\left[1-
  \left(\frac{T}{T_{c}}\right)^{\frac{2}{2b-1}}\Theta\left(
  b-\frac{1}{2}\right)\right],
\end{equation}
which can turn out to be more convenient for investigations.
$\Theta(x)$ is the Heaviside step function. Since
\begin {equation}\label{27}
  \Delta S(T)=-\frac{d}{dT}\Delta\Omega(T)\qquad {\rm{and}}\qquad
  \Delta C(T)=-T\frac{d^{2}}{dT^{2}}\Delta\Omega(T),
\end{equation}
where $\Delta S(T)$ and $\Delta C(T)$ are differences of the entropy
and the specific heat, respectively, differentiating the formula
(\ref{24}) we obtain
\begin {equation}\label{28}
  \Delta S(T)=\frac{2}{(2b-1)^{2}}N(n)T^{\frac{2}{2b-1}-1}
  \int_{T}^{T^{\ast}}
  \frac{\Delta^{2}(T^{\prime})}{(T^{\prime})^{\frac{2}{2b-1}+1}}
  dT^{\prime}
  - \frac{1}{2b-1}N(n)\frac{\Delta^{2}(T)}{T},
\end{equation}
which can be also written in an another quite equivalent form
\begin {equation}\label{29}
  \Delta S(T)=\frac{2}{(2b-1)^{2}}N(n)T^{\frac{2}{2b-1}-1}
  \int_{T}^{T^{\ast}}
  \frac{\Delta^{2}(T^{\prime})-\Delta^{2}(T)}{(T^{\prime})^
  {\frac{2}{2b-1}+1}}dT^{\prime}
  -\frac{1}{2b-1}N(n)\frac{\Delta^{2}(T)}
  {T_{c}}\left(\frac{T}{T_{c}}\right)^{\frac{2}{2b-1}-1}
  \Theta\left(b-\frac{1}{2}\right).
\end{equation}
Now, employing formula (\ref{28}) we can find the specific heat
difference in the form
\begin {equation}\label{30}
  \Delta C(T)=\frac{2(3-2b)}{(2b-1)^{3}}N(n)
  T^{\frac{2}{2b-1}-1}\int_{T}^{T^{\ast}}\frac{\Delta^{2}
  (T^{\prime})}{(T^{\prime})^{\frac{2}{2b-1}+1}}dT^{\prime}
  -\frac{3-2b}{(2b-1)^{2}}N(n)\frac{\Delta^{2}(T)}{T}
  -\frac{2}{2b-1}N(n)\Delta(T)\frac{d}{dT}\Delta(T),
\end{equation}
which can be also written in another quite equivalent form
\begin {eqnarray}\label{31}
  \Delta C(T)&=&\frac{2(3-2b)}{(2b-1)^{3}}N(n)
  T^{\frac{2}{2b-1}-1}\int_{T}
  ^{T^{\ast}}\frac{\Delta^{2}(T^{\prime})-\Delta^{2}(T)}{(T^{\prime})^
  {\frac{2}{2b-1}+1}}dT^{\prime}
  -\frac{3-2b}{(2b-1)^{2}}N(n)\frac{\Delta^{2}(T)}{T_{c}}
  \left(\frac{T}{T_{c}}\right)^{\frac{2}{2b-1}-1}
  \Theta\left(b-\frac{1}{2}\right)\nonumber\\
  &&-\frac{2}{2b-1}N(n)\Delta(T)\frac{d}{dT}\Delta(T).
\end{eqnarray} The obtained formulas include the parameter $b$ which is
an arbitrary positive number. Hence, if we could identify $b$ we would
be able to find the forms of the thermodynamic potential (free energy),
the entropy and the specific heat for the superconducting phase based
upon $\Delta(T)$ shape. Note that for BCS-like models $b=1$, and this
case has been considered by us in details before
\cite{pap25,pap26,pap27,pap28}. On the other hand, if we knew
$\Delta(T)$ and $\Delta C(T)$, then by virtue of the presented
formalism we could estimate $b$. We emphasize that in general $N(n)$,
i.e. the density of states defined above for the superconducting phase,
is different from that one defined for the normal phase. This
difference can be explained as a modification of the effective mass, and
must be included in all normal-phase thermodynamic functions when we
derive the thermodynamic functions for the superconducting phase from
the above formulas. This procedure is necessary in some approaches
\cite{pap25,pap29} to eliminate non-physical results, e.g. negative
values of the superconducting specific heat.
Note that the obtained Eqs.~(\ref{28})--(\ref{31}) imply a formal
relation between $\Delta S(T)$ and $\Delta C(T)$, which can be written
in the form
\begin {equation}\label{32}
  \Delta C(T)=\frac{3-2b}{2b-1}\Delta S(T) -\frac{1}{2b-1}N(n)
  \frac{d}{dT}\Delta^{2}(T).
\end{equation}

In order to consider the specific heat in low temperature limit
($T\rightarrow 0$), we have to take into account the normal specific
heat $C_{N}(T)$ which for this case in the frame of the presented
formalism can be found in the form
\begin{equation}\label{33}
  C_{N}(T)=T\int^{\infty}_{0}\bar{N}(2T u)\,u^{2}
  \cosh^{-2}(u)\,du,
\end{equation}
where
\begin{equation}\label{34}
  \bar{N}(2T u)=\frac{2}{\pi^{d-1}}\langle J(2T u,
  \vartheta,\varphi)+J(-2T u,\vartheta,\varphi)\rangle
\end{equation}
is another form of the density of states (cf. Eq.~(\ref{9})).
Since $\bar{N}(2T u)$ depends on $T$, $C_{N}(T)$ it can become
a complicated function of temperature.
According to Eq.~(\ref{32}) and including Eq.~(\ref{33}) the
superconducting specific heat can be expressed in the form
\begin {equation}\label{35}
  C_{S}(T)=\frac{3-2b}{2b-1}\Delta S(T)
  -\frac{2}{2b-1}N(n)\Delta(0)\frac{d}{dT}\Delta(T)+C_{N}(T).
\end{equation}
Hence, we state that the formula for the superconducting specific
heat can contain several different terms which determine the
temperature dependence. For the BCS model with a \textit{s-} or
\textit{d-}pairing the superconducting specific heat is proportional to
$d\,\Delta(T)/d\,T$ only, since for $b=1$ the term $\Delta S(T)$ is
proportional to $T$ and it is compensated by $C_{N}(T)$
\cite{pap14,pap25,pap28}.

Let us now consider the specific heat jump at the critical temperature
$T=T_{c}$. In order to do that comprehensively, we have to assume that
for $T\rightarrow T_{c}$ the energy gap
$\Delta(T)\rightarrow\Delta(T_{c})$ and $\Delta(T_{c})=0$
for the second-order (continuous) transition or $\Delta(T_{c})>0$ for
the first-order (discontinuous) transition, and for $T>T_{c}$ the
energy gap vanishes, i.e. $\Delta(T)\equiv 0$.
Then, the specific heat jump can be found from Eq.~(\ref{32})
in the limit $T\rightarrow T_{c}$ as
\begin {equation}\label{36}
  \Delta C(T_{c})=\frac{3-2b}{2b-1}\Delta S(T_{c})
  -\frac{2}{2b-1}N(n)\lim_{T\rightarrow T_{c}}
  \Delta(T)\frac{d}{dT}\Delta(T).
\end{equation}
Since for continuous phase transitions $\Delta S(T_{c})=0$, the
obtained formula coincides with that known for BCS-type models
\cite{pap18,pap19,pap30}.

%%%%%%%%%%%%%%%%%%%%%%%%%%%%%%%%%%%%%%%%%%%%%%%%%%%%%%%%%%%%%%%%%%%%%%%
\section{Reconstruction of the energy gap}
%%%%%%%%%%%%%%%%%%%%%%%%%%%%%%%%%%%%%%%%%%%%%%%%%%%%%%%%%%%%%%%%%%%%%%%

In the previous section we have shown that the thermodynamic potential,
the entropy and the specific heat differences can be defined based
upon the temperature dependence of the energy gap only. In this section
we show that also the energy gap $\Delta(T)$ can be reconstructed from
the specific heat difference if the values of $b$ and $\Delta S(0)$ are
known. Employing Eqs.~(\ref{27}) and (\ref{32}), after some
transformations, we can express the energy gap by means of
$\Delta C(T)$ and $\Delta S(0)$ in the form
\begin {equation}\label{37}
  \Delta^{2}(T)=\Delta^{2}(0)+\frac{3-2b}{N(n)}T\Delta S(0)
  +\frac{1}{N(n)}\int^{T}_{0}\left[(3-2b)\int^{T^{\prime}}_{0}
  \frac{\Delta C(T^{\prime\prime})}{T^{\prime\prime}}
  dT^{\prime\prime}-(2b-1)\Delta C(T^{\prime})\right]dT^{\prime}.
\end{equation}
Hence, for the cases when $C_{N}(T)\sim T^{q}$, where according to the
Nernst law $q>0$, Eq.~(\ref{37}) can be reduced to the form
\begin {eqnarray}\label{38}
  \Delta^{2}(T)&=&\Delta^{2}(0)+\frac{3-2b}{N(n)}T\Delta S(0)
  -\frac{3-2b-(2b-1)q}{N(n)q(q+1)}T C_{N}(T)\nonumber\\
  &&+\frac{1}{N(n)}\int^{T}_{0}\left[(3-2b)\int^{T^{\prime}}_{0}
  \frac{C_{S}(T^{\prime\prime})}{T^{\prime\prime}}dT^{\prime\prime}
  -(2b-1) C_{S}(T^{\prime})\right]dT^{\prime},
\end{eqnarray}
which for BCS-type models, when $\Delta S(0)=0$ and $b=q=1$,
includes $C_{S}(T)$ only \cite{pap26}.

%%%%%%%%%%%%%%%%%%%%%%%%%%%%%%%%%%%%%%%%%%%%%%%%%%%%%%%%%%%%%%%%%%%%%%%%%%%%
\section{Critical magnetic induction}
%%%%%%%%%%%%%%%%%%%%%%%%%%%%%%%%%%%%%%%%%%%%%%%%%%%%%%%%%%%%%%%%%%%%%%%%%%%%

The formulas obtained in Section \ref{unirel} allow us to find the
critical magnetic field $H_{c}(T)$ or the critical magnetic
induction $B_{c}(T)=\mu_{0}H_{c}(T)$ and their derivatives for
superconducting systems as functionals of $\Delta(T)$, only. In
accordance to the thermodynamic relation
\begin{equation}\label{38a}
  G_{S}(T,H=0)-G_{N}(T,H=0)=-\frac{1}{2}\mu_{0}H^{2}_{c}(T)
\end{equation}
after taking into account that
$G_{S}(T,H=0)-G_{N}(T,H=0)=\Delta\Omega(T)$
when the chemical potential is independent of temperature and
identical in both phases,  the critical magnetic induction can be
presented in the form
\begin{equation}\label{38b}
  B_{c}(T)=B_{c}(0)\sqrt{\frac{-2\Delta\Omega(T)}{N(n)\Delta^{2}(0)}},
\end{equation}
where
\begin{equation}\label{38c}
  B_{c}(0)=\sqrt{2\mu_{0}N(n)}\;\Delta(0),
\end{equation}
and $\Delta\Omega(T)$ should be taken in one of the forms
(\ref{26})--(\ref{28}). In a similar way employing standard
relations and including forms of Eqs.~(\ref{30})--(\ref{33}) we can
express derivatives of the critical magnetic induction in the
form
\begin{eqnarray}
  \frac{dB_{c}(T)}{dT} &=&\sqrt{\frac{\mu_{0}}{2}}\;\frac{\Delta S(T)}{\sqrt{-
  \Delta\Omega(T)}},\label{38d}\\
  \frac{d^{2}B_{c}(T)}{d^{2}T}&=&\sqrt{\frac{\mu_{0}}{2}}
  \left[\frac{\Delta C(T)}{T\sqrt{-\Delta\Omega(T)}}
  -\frac{(\Delta S(T))^{2}}{(-\Delta\Omega(T))^{\frac{3}{2}}}\right],\label{38e}
\end{eqnarray}
which should be useful to estimate values of the introduced
parameter $b$ for some unconventional superconductors by means of
experimental data.

%%%%%%%%%%%%%%%%%%%%%%%%%%%%%%%%%%%%%%%%%%%%%%%%%%%%%%%%%%%%%%%%%%%%%%%
\section{Remarks on McMillan's case}
%%%%%%%%%%%%%%%%%%%%%%%%%%%%%%%%%%%%%%%%%%%%%%%%%%%%%%%%%%%%%%%%%%%%%%%

In the formalism developed by McMillan \cite{pap24}, where
strong-coupling electron-phonon effects were included, the critical
temperature was derived in the form
\begin{equation}\label{39}
  T_{c}\sim\omega_{D}\exp\left(-\frac{1+\lambda}{\lambda-\mu^{\ast}
  (1+\lambda)}\right),
\end{equation}
where the parameter $\mu^{\ast}$ expresses the Coulomb repulsion
between electrons forming a Cooper pair, and for known materials
$\mu^{\ast}\simeq 0.13$, $\omega_{D}$ is the Debye energy which is the
actual cut-off parameter for conventional superconductors.
Note that the critical temperature $T_{c}=0$ for $\lambda=
\frac{\mu^{\ast}}{1-\mu^{\ast}}>0$, so in obedience to the presented
consideration we should rather use an effective coupling constant which
varies form $0$ to its real value when the coupling interaction is
included. Such effective coupling constant is
$\lambda_{\rm{eff}}=a^{-1}[\lambda-\mu^{\ast}(1+\lambda)]$. Hence,
putting it into Eq.~(\ref{39}) we obtain \cite{pap14}
\begin{equation}\label{40}
  T_{c}\sim\omega_{D}\exp(-\frac{1}{1-\mu^{\ast}})\exp\left(-\frac{b}
  {\lambda_{\rm{eff}}}\right),
\end{equation}
where $b^{-1}=a(1-\mu^{\ast})>0$ and $a$ is positive constant which
ensures that $\Delta(0)$ keeps the form (\ref{17}). Hence, the
formalism developed by us can be applied to the McMillan-type
superconductors.

%%%%%%%%%%%%%%%%%%%%%%%%%%%%%%%%%%%%%%%%%%%%%%%%%%%%%%%%%%%%%%%%%%%%%%%%%%%%
\section{Application of derived relations to high-$T_{c}$ superconductors}
%%%%%%%%%%%%%%%%%%%%%%%%%%%%%%%%%%%%%%%%%%%%%%%%%%%%%%%%%%%%%%%%%%%%%%%%%%%%%

The mechanism of superconductivity for HTSC based on
antiferromagnetic correlations was firstly proposed to predict
antiferromagnetic behaviour for pure $\rm{La_{2}CuO_{4}}$. As soon as
these materials are doped, giving a few per cent of holes in the
$\rm{CuO_{2}}$ plane, the antiferomagnetism is destroyed. As the number
of carriers $\delta$ is always small, they cannot
completely screen the Coulomb interaction on copper site. Thus the copper
electrons can move but are always subject to strong on-site
repulsion. The introduction of additional holes causes that in
sufficiently low temperatures they are coupled by superexchange
interaction, through the oxygen atoms, forming Cooper pairs
\protect\cite{pap31,pap32,pap33,pap34,pap35,pap36}.  The similar
situation occurs in the compound $\rm{YBa_{2}Cu_{3}O_{6}}$ which is
an insulating antiferromagnetic. In this so-called {\textit{t--J}} model the
gap equation can be obtained in the form (\ref{1a})
by performing a BCS linearisation of the $J$ term. Then it can be expressed
in the form \protect\cite{pap24a}
\begin{equation}\label{41}
  \frac{2}{J}=\frac{1}{N}\sum_{\bf k}\gamma^{2}_{\bf k}
  \frac{\tanh(\beta E_{\bf k}/2)}{E_{\bf k}},
\end{equation}
where
\begin{equation}\label{42}
  \gamma_{\bf k}=\cos(k_{x}a)\pm\cos(k_{y}a)
\end{equation}
and $\pm$ refer to $s$-wave and $d$-wave superconductivity,
respectively. Note that $\gamma_{\bf k}$ represents a factorable
pairing interaction which determines the form of the energy gap.
So, it is compatible with the form of the pairing interaction assumed
in the presented approach. Moreover, Cyrot et al.
\protect\cite{pap24a} in order to derive the critical temperature
introduced some restrictions on the form of the density of states
separating in the integral form of the gap equation (\ref{41}) with
$\Delta=0$ two domains of integration: the first one where the
density of states can be approximated by a constant value, and the
second one where the hyperbolic tangent can be replaced by $1$.
Declaring that their solution is valid only for the range of
doping $\delta\geq 0.05$, when the order parameter has the $s$
symmetry, they obtained the critical temperature in the form
\begin{equation}\label{43}
  T_{c}=1.14W(\mu)\exp[-1/JN(\mu)]
\end{equation}
where $JN(\mu)$ corresponds to the dimensionless coupling constant
$\lambda$, and $W(\mu)$ is proportional to $\omega_{c}$
(cf. Eq.~(\ref{13}) and Eq.~(\ref{17})). Moreover, they showed that
at zero temperature $\gamma_{\bf k}(\mu) J\Delta(0)=1.76 T_{c}$. So,
the {\textit{t--J}} model within the developed approach
\protect\cite{pap24a} reveals BCS-like properties with $b=1$.

In order to employ the found universal relations to the {\textit{t--J}}
model we have to put $b=1$, since the other quantities or
representing them expressions are not involved in these relations.
Moreover, let us emphasize that in our method the summation over ${\bf
k}$ is replaced by the integration over $\xi=\epsilon -\mu$ for an in
general non-symmetric pairing interaction by virtue of the
conformal transformation, in which the Jacobian represents the
scalar field of the density of states \protect\cite{pap20}. It
causes that our method is more precise than the one discussed above, and
it allows us to predict that $b$ can be different than $1$.
Note that the three substantial assumptions mentioned above imply the
following facts: the constancy of the density of states corresponds
to putting the Jacobian constant, the inclusion of a symmetric
pairing interaction is equivalent to $n=1$, and for $\omega\equiv\omega_D$,
same as the hyperbolic tangent, the function $f$ can be replaced
by 1. As a consequence the parameter $b$ given by Eq.~(\ref{14}) reduces
to 1.

%%%%%%%%%%%%%%%%%%%%%%%%%%%%%%%%%%%%%%%%%%%%%%%%%%%%%%%%%%%%%%%%%%%%%%%
\section{Conclusions}
%%%%%%%%%%%%%%%%%%%%%%%%%%%%%%%%%%%%%%%%%%%%%%%%%%%%%%%%%%%%%%%%%%%%%%%

The presented formalism proves that there exist some universal
relations between the free energy, entropy, specific heat
differences, and the energy gap amplitude for a wide class of
models of superconductivity. In our previous papers
\cite{pap25,pap27,pap28} we have shown that applying the formulae
(\ref{30}) or (\ref{31}) established for the case when $b=1$
towards a general \textit{s-}paired BCS case, when the magnetic
field and superflow are taken into account, and for a pure
\textit{d-}paired BCS model, we obtained correct forms of the
superconducting specific heat, where the linear (with respect to
temperature) terms were entirely eliminated by $C_{N}(T)$.
Therefore we expect now that these relations will be verified
experimentally, and they will turn out to be helpful in
explanation of common elements of the mechanism of low- and
high-$T_{c}$ superconductivity. On the other hand, we remind that
we imposed some restrictions on conditions which must be satisfied
by the superconducting system. Hence, the presented formalism can
be applied to superconducting systems when the cut-off parameter $\omega_D$
exceeds $\Delta(0)$ or $v_{l}\Delta(0)$ by far, the chemical
potential is identical in both phases, and the pairing interaction
can be taken in a form corresponding to BCS-type models, which can
be always attained by performing a BCS linearisation of
the appropriate interaction term.

%%%%%%%%%%%%%%%%%%%%%%%%%%%%%%%%%%%%%%%%%%%%%%%%%%%%%%%%%%%%%%%%%%%%%%%
% References
%%%%%%%%%%%%%%%%%%%%%%%%%%%%%%%%%%%%%%%%%%%%%%%%%%%%%%%%%%%%%%%%%%%%%%%

\end{document}